\documentclass[12pt]{iopart}

\usepackage{iopams}  
\usepackage{color}
\usepackage{cite}
\usepackage{graphicx,graphics}
\usepackage{mathrsfs}
\usepackage{multirow}
\usepackage{fixdif} 
\def\bra#1{{\left\langle #1 \right|}}
\def\ket#1{{\left| #1 \right\rangle}}
\def\braket#1{{\left\langle #1 \right\rangle}}
\providecommand{\proj}[1]{|#1\rangle\langle#1|}

\renewcommand{\tr}[1]{{\sf Tr}\left[#1\right]}

\providecommand{\bL}{\boldsymbol{\Lambda}}
\providecommand{\bU}{\boldsymbol{U}}
\providecommand{\bV}{\boldsymbol{V}}

\providecommand{\Ntot}{\hat N_{\sf tot}}
\providecommand{\aop}{\hat a}
\providecommand{\adag}{\hat a^\dagger}
\providecommand{\sop}{\hat s}
\providecommand{\sij}[1]{\hat s_{#1}}

\providecommand{\Sij}[1]{\hat S_{#1}}

\providecommand{\pop}{\hat{p}}

\providecommand{\Aop}{\hat{A}}
\graphicspath{ {./images/} }

\begin{document}

\title[Multilevel Quantum Rabi Models]{Multilevel Quantum Rabi Models}

\author{Tabitha Doicin}
\address{School of Mathematical Sciences, University of Nottingham, University Park, Nottingham NG7 2RD, UK}
\author{Andrew D. Armour}
\address{School of Physics and Astronomy and Centre for the Mathematics and Theoretical Physics of Quantum Non-Equilibrium Systems, University of Nottingham, Nottingham NG7 2RD, UK}
\author{Tommaso Tufarelli}
\address{School of Mathematical Sciences, University of Nottingham, University Park, Nottingham NG7 2RD, UK}
\vspace{10pt}

\begin{abstract}
The quantum Rabi model, which  describes the interaction between a simplified atom and a mode of the electromagnetic field, is a cornerstone of modern quantum optics. One of the key assumptions of the model is that the `atom' is a perfect two-level system. We explore what happens when one generalizes the atom to a multilevel system, with $m$ ground and $n$ excited states coupled to the same field. We focus on the case where the excited and ground states form distinct, well-separated, manifolds of near-degenerate levels (so that the spacing between the excited states on a given manifold is much less than the average spacing between the manifolds) and consider either uniform or random couplings between the individual ground and excited levels. We find that the system reduces approximately to a direct sum of Rabi models with a range of different couplings. Importantly, the strongest coupling is enhanced in a way that depends on the number of levels, for the simple case where $n=m$, the coupling scales with $n$ for uniform couplings and $2\sqrt{n}$ for random couplings (in the limit of large $n$). Our work thus suggests that multilevel Rabi systems could provide an attractive alternative route to accessing regimes of very strong coupling in light-matter systems. 
\end{abstract}

%
%
%
%
%

\section{Introduction}
The quantum Rabi model (QRM) is the simplest fully quantum mechanical model of light-matter interaction. It comprises a 2-level atom interacting with a single mode field. Although originally described decades ago\,\cite{Jaynes1963}, an analytic solution was only obtained in 2011\,\cite{Braak2011} and the QRM remains a topic of very active research\,\cite{Twyeffort2007,Braak2017,Xie2017,Rossatto2017,Solano2019,Twyeffort2022}. A key focus of work on the QRM has been the exploration of regimes where the atom and field become very strongly coupled\,\cite{ Twyeffort2007,Braak2017,Xie2017,Rossatto2017,Solano2019,Kockum2019}, so that a rotating wave approximation (leading to the much simpler Jaynes-Cummings model\,\cite{Jaynes1963}) is no longer valid. Recent work has uncovered the richness of the physics that the QRM can display, including entangled ground states with non-zero photon populations, as well as identifying a range of possible applications in quantum technology\,\cite{Solano2019,Kockum2019}. Furthermore, there has also been very rapid experimental progress in engineering quantum devices in which the QRM model can be realised with couplings many orders of magnitude larger than those which can be achieved with more traditional atomic systems\,\cite{Solano2019,Kockum2019}.

Strongly enhanced couplings can also be achieved in systems described by generalisations of the Rabi model. The best known example is the Dicke model\,\cite{Kockum2019}, where a single mode interacts with $n$ identical two-level atoms, leading to an effective interaction strength that scales as $\sqrt{n}$. Very recently, it was shown that a similar $\sqrt{n}$ enhancement can be achieved in a version of the Jaynes-Cummings model where a mode couples a single atomic ground state to a set of $n$ almost-degenerate excited states when the coupling strength is uniform\,\cite{Tufarelli_2021}. This model can be seen as an interesting theoretical generalisation, but it also has practical significance: it has been used to explain the observation of strong coupling effects in experiments involving colloidal quantum dots and a plasmonic nanoresonator\,\cite{Gross2018}.

In this paper we take a further step, generalising to consider a multilevel Rabi model in which a single field mode couples a set of $m$ nearly degenerate atomic ground states to a set of $n$ nearly degenerate excited states. Using a series of simple examples, we explore in detail the connection between different specific forms of the multilevel model and the standard QRM. In the simplest cases, where degeneracy exists for both the sets of excited and ground states, the system reduces to a direct sum of standard two-level Rabi models which can be labelled by the eigenvalues of a new symmetry operator. We find that the couplings within the individual Rabi models in general depend sensitively on the specific way in which the levels are coupled to the field. Turning to larger systems and focusing on the case where $n=m$, we investigate the scaling of the strongest coupling with $n$. For a uniform matrix of couplings between each of the different ground and excited states, the scaling goes as $n$. The case of random couplings is much more involved, but using results from random matrix theory we obtain a detailed expression that can describe the scaling for $n\ge 2$ and which tends asymptotically to $2\sqrt{n}$ in the large-$n$ limit. 
We also show that, as one might expect, small detunings between the levels generally have a small impact on the spectrum of the system, except for isolated regions of the parameter space where the small detunings induce avoided crossings. 

The rest of this paper is organised as follows. In Sec.\ \ref{sec:MQRMmodel} we introduce the multilevel generalisation of the Rabi model and define the notation. Next, in  Sec.\ \ref{sec:FLE}, we illustrate the phenomenology of the multilevel system and its connection to the original two-level Rabi model through a series of simple examples containing just a pair of excited and grounds states ($n=m=2$). Then, in Sec.\ \ref{sec:MLS}, we consider much larger systems, focusing in particular on the scaling of the maximum light-matter couplings with the number of levels. Finally, we summarise, draw our conclusions and discuss interesting avenues for future work in Sec.\ \ref{sec:CCs}.







\section{Multilevel Rabi Model}
\label{sec:MQRMmodel}
We consider a single field mode with annihilation operator $\aop$ and bare frequency $\omega$ coupled to a multilevel atom with $n$ excited states $\ket{e_1},...,\ket{e_n}$, with bare energies $\omega +\varepsilon\delta_{e_1},...,\omega +\varepsilon\delta_{e_n}$ and $m$ ground states $\ket{g_1},...,\ket{g_m}$ with bare energies $\varepsilon\delta_{g_1},...,\varepsilon\delta_{g_m}$. For concreteness, we will take $n\ge m$, bearing in mind that the case $m>n$ is similar. We assume that $\delta_{e,i},\delta_{g,i}\in[-1,1]$, so that $\varepsilon$ sets the scale of the overall spread of the excited and ground states about $\omega$ and $0,$ respectively.  In the cases we will consider, the model is assumed to be either degenerate ($\varepsilon=0$) or nearly so, with $\varepsilon/\omega\ll 1$. However, we expect that many of the results we obtain in this paper can be extended beyond this simplified scenario. A complex matrix $\boldsymbol{\Lambda}$ with elements $\Lambda_{ij}$ specifies how each ground-to-excited atomic transition couples to the field quadrature $\aop+\adag$. We note that $\bL$ does not, in general, have any symmetries (note that it is not even a square matrix for $n\neq m$). The corresponding Hamiltonian takes the form
\begin{eqnarray}
    \hat H=\omega\Ntot+\varepsilon\sum_{i=1}^n\delta_{e_j}\proj{e_j}+\varepsilon\sum_{j=1}^m\delta_{g_j}\proj{g_j}
    \nonumber\\+\sum_{j=1}^m\sum_{i=1}^n\left(\Lambda_{ij}\sij{ji}+\Lambda_{ij}^*\sij{ji}^\dagger\right)\left(\hat a+\hat a^\dagger\right),
    \label{MQRMhamiltonian}
\end{eqnarray}
where we have introduced the notation
\begin{equation}
    \sop_{ji}\equiv\ket{g_j}\bra{e_i}
\end{equation}
and defined the total excitation number as
\begin{equation}
    \hat N_{\sf tot}=\hat a^\dagger \hat a+\sum_{i=1}^n\proj{e_i}.
\end{equation}
Hence we can see that the model retains the parity symmetry of the QRM: $[\hat{\Pi},\hat{H}]=0$, with
\begin{equation}
 \hat{\Pi}={\mathrm{e}}^{i\pi\hat{N}_{\sf tot}}   
\end{equation}
the parity operator.
The standard 2-level QRM is, of course, recovered if we set $n=m=1$.

\section{Few-Level Examples}
\label{sec:FLE}
In this section we illustrate the properties of the multilevel model and its relationship to the standard QRM using the simple case of 4 atomic levels ($m=n=2$) with ground states $\ket{g_1},\ket{g_2}$ and excited states $\ket{e_1},\ket{e_2}$. Starting with the case of degenerate levels ($\varepsilon=0$), and a diagonal coupling matrix $\bL = {\rm{diag}}(\lambda_1,\lambda_2)$, we explore how simply adding additional resonant atomic levels to the QRM can lead to a symmetry which we call `doublet symmetry'. Although this symmetry is not exact in cases where $\varepsilon\neq0$, the concept remains useful in analysing more complex scenarios. We then consider the case where the coupling matrix $\bL$ is not strictly diagonal, introducing the use of the radiation basis\,\cite{Tufarelli_2021} to simplify the model and demonstrating how the effective light-matter coupling is boosted. In the final subsection, we introduce small atomic detunings ($\varepsilon\neq0$) and investigate the symmetry breaking they generate.

\subsection{Degenerate Levels with Diagonal Couplings}
\label{sec:PDS}
We start with an extension of the standard QRM involving degeneracy in both the ground and excited states $(\varepsilon=0)$ together with a diagonal coupling matrix, $\bL={\sf diag}(\lambda_1,\lambda_2)$. The
 Hamiltonian simplifies to
\begin{equation}
    \hat H= \omega \Ntot + \sum_{j=1}^2 \lambda_j \left( \sij{jj}+\sij{jj}^\dagger \right)\left( \hat a+\hat a^\dagger \right),\label{Hrabi-2indep}
\end{equation}
where we can assume $\lambda_j\ge0$ without loss of generality, since it is always possible to redefine the phase of the states $\ket{e_j}$. 

\begin{figure}[t]
    \centering
    \includegraphics[width=0.2\linewidth]{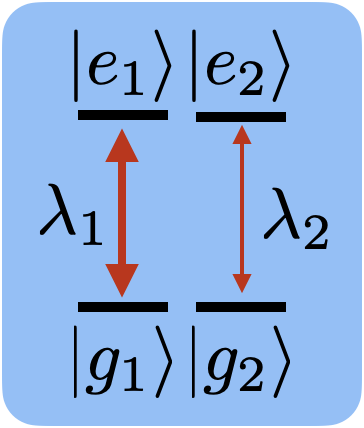}   
    \includegraphics[width=0.35\linewidth]{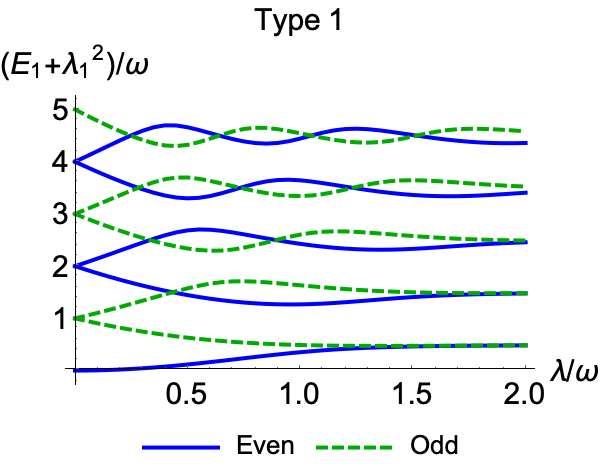}    \includegraphics[width=0.35\linewidth]{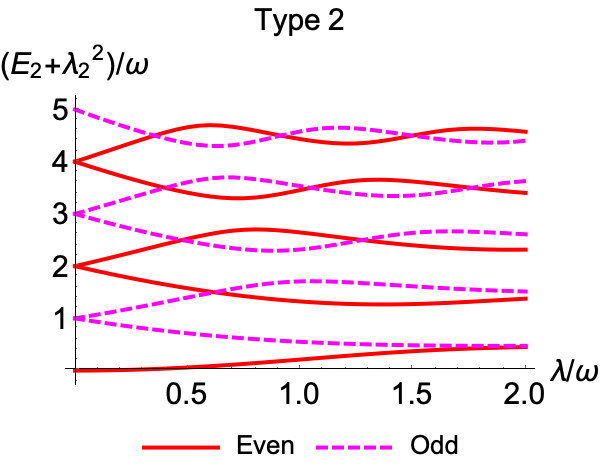}
    \caption{Left: Schematic of a four-level Rabi model with two degenerate excited states and two degenerate ground states. Each transition $\ket e_j\leftrightarrow \ket g_j$, couples to the field with strength $\lambda_j$, while transitions $\ket{e_i}\leftrightarrow \ket{g_j}$ with $i\neq j$ are not coupled.
    Centre and Right: First few energy eigenvalues of the model as a function of the coupling constant $\lambda$, fixing $\lambda_1=\lambda, \lambda_2=0.7\lambda$. The term $\lambda_{i}^2/\omega$ has been added to all energies for ease of visualization. Continuous (dashed) lines correspond to even (odd) parity  eigenstates.}
    \label{fig:IndependentRabis}
\end{figure}

In addition to the parity symmetry, this special case has an additional  `doublet symmetry'. The Hamiltonian commutes with the `doublet type' operator
\begin{equation}
    \hat D =\sum_{j=1}^2j \left(\proj{e_j}+\proj{g_j}\right). \label{eq:type}
\end{equation}
The operator $\hat D$ simply distinguishes the two possible `types' of states, by which we mean states labelled with different values of the index ($j=1,2$) of the atomic states: for example, the subspace $\{\hat D=1\}$ is spanned by  type-1 states of the form $\ket{\Psi_1}=\ket{e_1}\ket{\psi_e}+\ket{g_1}\ket{\psi_g},$ where $\ket{\psi_{e}},\ket{\psi_{g}}$ are arbitrary field states. The case $\{\hat D=2\}$ is analogous. The idea behind doublet symmetry is very simple: the Hamiltonian in Eq.~(\ref{Hrabi-2indep}) does not include any process that can mix the doublet of type-1 states with the doublet of type-2 state. Hence, this model is just a \textit{direct sum} of two Rabi models that are decoupled from each other\,\cite{Morris_1983}. The matrix representation of the Hamiltonian has the block form
\begin{equation}
    \hat{H}=\left(
    \begin{array}{cc}
        \hat{H}_1 & 0 \\
        0 & \hat{H}_2
    \end{array}\right),
\end{equation}
where the blocks $\hat H_{i}$, which only contain terms involving $\ket{g_i}, \ket{e_i}$, can be further decomposed into sub-blocks of even and odd parity. In summary, this example is readily understood in terms of the properties of the standard 2-level QRM. As illustrated in Fig.\ \ref{fig:IndependentRabis}, there are 2 families of energy eigenvalues corresponding to doublets of type (index) 1 or 2, each of which can be further subdivided into sub-families with even or odd parity.

\subsection{Non-Diagonal Couplings and the Radiation Basis}
We now consider the case where the off-diagonal terms of the coupling matrix are non-zero, such that for some $b\in(0,1]$

\begin{equation}
\bL = \lambda\left(
\begin{array}{cc}
1 & b \\
b & 1
\end{array}
\right)
\end{equation}
In this situation, it is helpful to switch from the original bare basis to a radiation basis\,\cite{Tufarelli_2021,Morris_1983}. The radiation basis is defined by using the singular value decomposition (SVD) of the coupling matrix to define a new atomic basis, together with two effective coupling constants. This allows us to view the model directly from the point of view of the interaction with the field mode: The change of basis groups the collective atomic modes that are either strongly or weakly coupled to the field. In particular, it reveals how specific superpositions of atomic states (rather than the bare states themselves) take part in the light-matter interaction. 

\begin{figure}[t]
    \centering
    \includegraphics[width=0.7\linewidth]{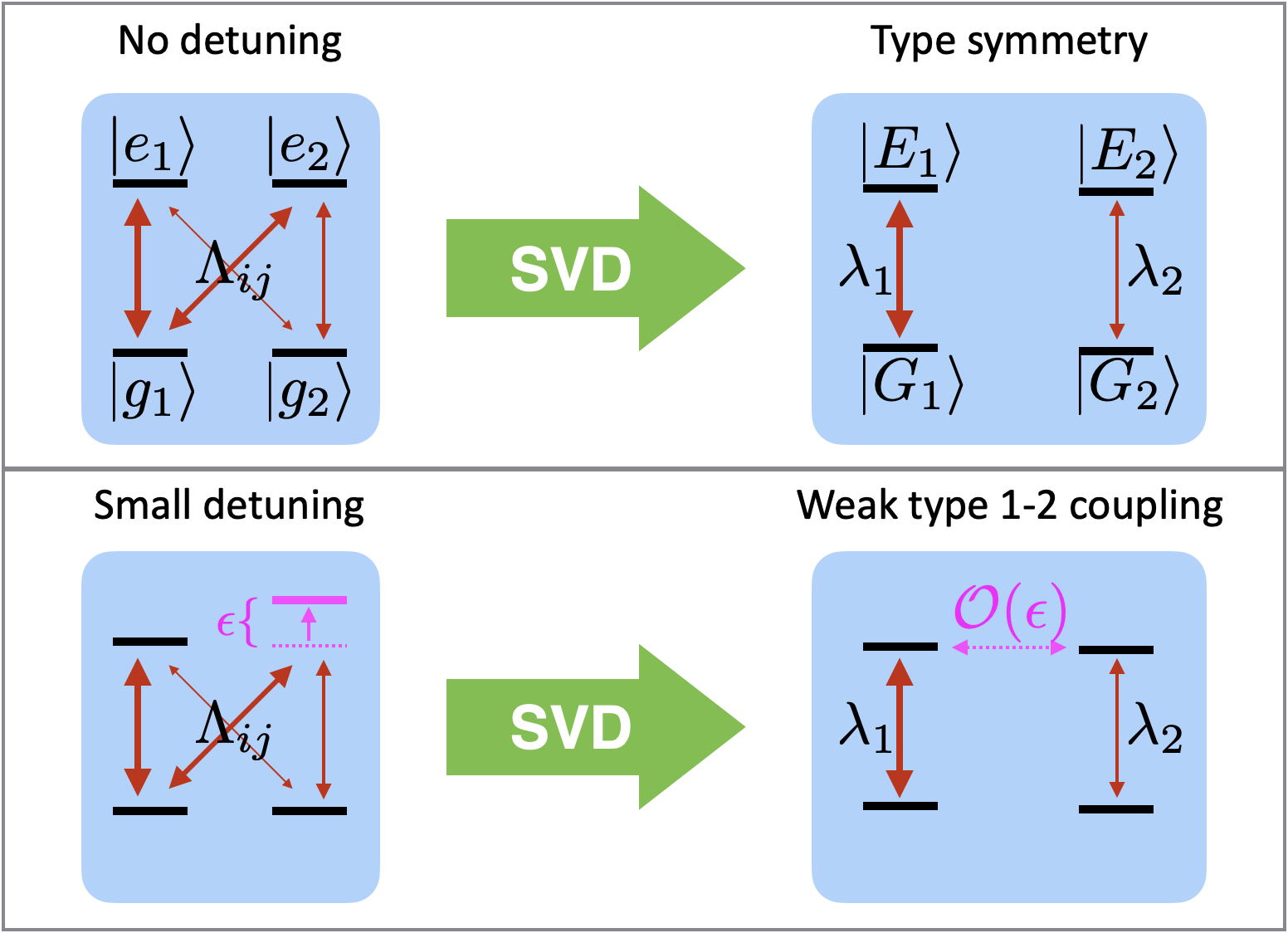}
    \caption{Schematic illustrations of 4-level Rabi models with a non-diagonal coupling. The upper panel shows the case where the excited and ground state manifolds are each degenerate (no detunings): a singular value decomposition (SVD) leads to a pair of uncoupled Rabi models in the radiation basis with different couplings distinguished by different eigenvalues of the doublet operator. The lower panel shows the case where one excited state is detuned by $\varepsilon\neq 0$: coupling between the doublets breaks the symmetry.}
    \label{fig:4levSVD}
\end{figure}

The SVD is described in detail in \ref{app:SVD}, the corresponding transformation to the radiation basis is given by
\begin{eqnarray}
\ket{g_1} = \frac{1}{\sqrt{2}}(\ket{G_1} + \ket{G_2}), \ket{g_2} =\frac{1}{\sqrt{2}}(\ket{G_1} - \ket{G_2}), \nonumber\\ \ket{e_1} = \frac{1}{\sqrt{2}}(\ket{E_1} + \ket{E_2}), \ket{e_2} =\frac{1}{\sqrt{2}}(\ket{E_1} - \ket{E_2}), 
\end{eqnarray} 
and the eigenvalues of $\bL$ are $\lambda_1=\lambda(1+ b),\lambda_2=\lambda(1- b)$. The interaction part of the Hamiltonian reduces from 4 terms to 2,  
\begin{equation} 
\hat{H}_{int}= \lambda_1\left(\ket{G_1}\bra{E_1} + \ket{E_1}\bra{G_1}\right) + \lambda_2\left(\ket{G_2}\bra{E_2} + \ket{E_2}\bra{G_2}\right). 
\label{simpleexamplesinteraction}
\end{equation}
This is now structurally the same as the case described in the previous subsection\,\cite{Morris_1983,Zlatonov_2020}: a direct sum of two independent 2-level Rabi models with effective couplings $\lambda_{1(2)}$, as illustrated in the upper panel of Fig.\ \ref{fig:4levSVD}. Note that the system retains a doublet symmetry whereby the appropriate doublet operator (Eq.\ \ref{eq:svdtype} in \ref{app:SVD}) commutes with the Hamiltonian. Note that we adopt the convention that the lower the type index, the higher the corresponding effective coupling.

As we vary $b$, the couplings between different atomic subsystems and light change. At its maximum ($b=1$), the type-1 ($\braket{D}=1$) doublet (states $\ket{G_1},\ket{E_1}$), display a doubly enhanced light-matter coupling compared to the bare basis states. In contrast, the type-2 states, $\ket{G_2}$ and $\ket{E_2}$, become dark states: they become completely decoupled from the field mode. This coupling enhancement is very similar to superradiant phenomena in collective light-matter models such as the Dicke model\,\cite{DickeOriginal,Dicke_modern,Tufarelli_2021}, for which certain superpositions of states become strongly coupled, while others decouple entirely.

\subsection{Atomic Detunings}

So far, our analysis has relied on perfect degeneracy in the atomic levels ($\varepsilon = 0$), which allowed a clean separation into independent doublets. If this perfect degeneracy is broken, ($\varepsilon\neq0$), the clean separation into doublets (labelled by type) is lost. Parity symmetry will always remain intact, but the radiation basis would no longer diagonalise the entire atomic structure, leading to mixing between doublet types. We illustrate this using a simple example, where we again assume $n=m=2$, but now set $0<\varepsilon\ll\omega$.

\begin{figure}[t]
    \centering
    \includegraphics[width=0.6\linewidth]{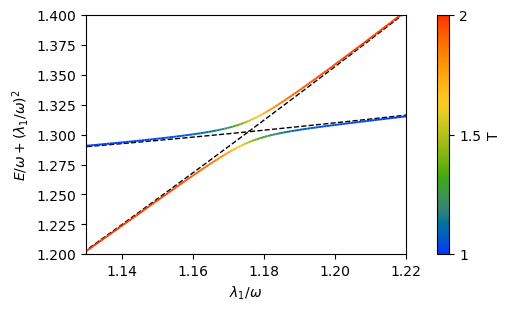}
    \caption{Energy level diagram showing an example of the level repulsion which occurs when detunings are nonzero. Full lines show the anti-crossing of two of the eigenergies of the 4-level system for $\varepsilon=0.015\omega$, $\delta_{e/g1}=-1, \delta_{e/g2}=1$, whilst dashed lines show the case where there is no detuning ($\varepsilon=0$). The colour of the anticrossing eigenvalues is coded by the corresponding expectation value of the doublet operator, $\hat{D}$, illustrating the rapid transition in doublet type that occurs as the system evolves through the avoided crossing.}
    \label{fig:DoubletMixing}
\end{figure}

We can follow the same steps as before: collect all the couplings into the matrix $\bL$, and then use the singular value decomposition (SVD) of the matrix to define a new atomic basis and two effective coupling constants for this basis, $\lambda_1$ and $\lambda_2$. The interaction part of the Hamiltonian is therefore the same as in the previous subsection, [Eq. \ref{simpleexamplesinteraction}]. However, the atomic detuning terms introduce new structure when expressed in the radiation basis, leading to off-diagonal couplings between the doublet types. These terms transform as
\begin{equation}
    \hat H_\varepsilon=\varepsilon\sum_{i=1}^n\delta_{e_j}\proj{e_j}+\varepsilon\sum_{j=1}^m\delta_{g_j}\proj{g_j} = \hat{H}_{\varepsilon}^e + \hat{H}_{\varepsilon}^g,\label{Hrabi-2indep2}
\end{equation}
where 
\begin{eqnarray}\fl
        \hat{H}_{\varepsilon}^e  =\frac{\varepsilon(\delta_{e1} + \delta_{e2})}{2}(\ket{E_1}\bra{E_1} + \ket{E_2}\bra{E_2}) + \frac{\varepsilon(\delta_{e1} - \delta_{e2})}{2}(\ket{E_1}\bra{E_2} + \ket{E_2}\bra{E_1}),\nonumber\\
        \fl
        \hat{H}_{\varepsilon}^g = \frac{\varepsilon(\delta_{g1} + \delta_{g2})}{2}(\ket{G_1}\bra{G_1} + \ket{G_2}\bra{G_2}) + \frac{\varepsilon(\delta_{g1} - \delta_{g2})}{2}(\ket{G_1}\bra{G_2} + \ket{G_2}\bra{G_1}). 
\end{eqnarray}
As before, the full Hamiltonian includes two Rabi models that we could call again type 1 and type 2 (described by Eq.\ \ref{simpleexamplesinteraction}). However, they are no longer independent: the detuning, $\varepsilon$, generates a weak coupling between the two models that mixes the type of the excited state. Recalling the block-structure picture from Sec.\ \ref{sec:PDS}, the Hamiltonian now gains weak off-diagonal terms linking the two previously independent blocks. Although the doublet symmetry is no longer exact, it remains a valuable approximate symmetry when $\varepsilon \ll \omega$. In this regime it is clear that most eigenstates can still be meaningfully classified by their dominant doublet character. The key exceptions arise near avoided crossings, where the energy levels of different types approach and strongly hybridise due to the detuning-induced coupling. Since $\varepsilon$ sets the size of the anti-crossing it can no longer be considered small and indeed the type of the eigenvalues swaps over as illustrated in Fig.\ \ref{fig:DoubletMixing}. 


Finally, we note that when $m\neq n$, there are residual atomic states that form additional dark states arise with no coupling to the field (see \ref{app:SVD}). At first sight these seem irrelevant, but for $\varepsilon\neq 0$ weak couplings between states in the radiation basis mean that even the dark states generally participate in (and complicate) the dynamics.

In summary, while detunings break the exact doublet symmetry, they do so in a very controlled way if the detunings are small. The approximate symmetry continues to structure the spectrum in a way that provides a useful framework for analytical and numerical approaches.
\section{Coupling Scalings in Multilevel Systems}
\label{sec:MLS}

We now move on to consider larger systems, focusing in particular on systems with either uniform or random couplings. We start by looking at how the coupling can be boosted for the uniform and random cases, i.e. how the strongest coupling (after the SVD has been performed) scales with the number of near-degenerate levels, before going on to look at the overall structure of the energy level distribution for random couplings.

\subsection{Coupling Boost for Uniform Couplings}
The situation where all of the individual transitions have the same coupling is an important (and simple) special case. As we have already seen, for $n=m=2$, the SVD yields the diagonal elements $\lambda_1=2\lambda,\;\lambda_2=0$, so that the system features a single nontrivial Rabi model. It is easy to generalize this result to $n=m>2$ via a constant coupling matrix $\Lambda_{ij}=\lambda$, which yields  $\lambda_1=n\lambda,$ $\lambda_{j>1}=0,$ i.e. the only nontrivial Rabi model embedded in the system has a coupling constant that grows linearly in the number of states. This is in contrast to the case where there is only a single ground state ($m=1$), but multiple excited states\,\cite{Tufarelli_2021} ($n>1$) , where the coupling scales $\sim\sqrt{n}$. 

Interestingly, the reduction of the multilevel system to a single effective Rabi model extends beyond the case of purely uniform coupling. If it is possible to factorize the coupling matrix in terms of two vectors ${\bf{v}},{\bf{w}}$, so that $\Lambda_{ij}=v_iw_j$ then one finds $\lambda_1=|\mathbf{v}||\mathbf{w}|,\lambda_{j>1}=0.$

\subsection{Coupling Boost for Random Coupling Matrix} \label{random couplings section}

The case of a random coupling matrix is the opposite extreme to that of uniform couplings. Our goal is to find an approximate distribution for the largest singular value, $\lambda_1$, of our complex $n\times m$ coupling matrix $\bL$, so that we can analyse its statistical properties. To do this, we employ tools from random matrix theory and leverage known results on the eigenvalue distribution of Wishart ensemble matrices.  We leave most of the details of the calculation to \ref{app:stats}, and simply focus on the main results here. 

We assume a coupling matrix, $\bL$, with elements that are independent and identically distributed (IID), taken from a complex Gaussian distribution with mean 0 and variance 1, denoted $\mathscr{CN}(0,1)$. A random variable is said to follow a complex Gaussian distribution $Z \sim \mathscr{CN}(0,1)$ if $Z=X+iY$, where $X$ and $Y$ are IID real Gaussian random variables $X,Y\sim \mathscr{N}(0,1/2)$. The coupling matrix $\bL$ would then be said to belong to the Ginibre ensemble class of matrices\,\cite{ginibreensemblesreview}.

Rather than analyzing the singular value distribution of the coupling matrix directly, we use the result that the largest singular value of $\bL$ corresponds to the largest eigenvalue $\kappa_1$ of the associated Wishart matrix $W = \bL \bL^T$. Since the distribution of $\kappa_1$ is well-approximated by a known form\,\cite{CHIANI_RMT}, this approach provides a practical and accurate way to characterize $\lambda_1$. Since the singular values of $\bL$ are defined as the square roots of the eigenvalues of $W = \bL \bL^T$, we have:
\begin{equation}
    \lambda_1 = \sqrt{\kappa_1}.
    \label{sqrt}
\end{equation}
This transformation allows us to approximate the distribution of $\lambda_1$ using the well-characterized properties of $\kappa_1$.

The approximate distribution for the largest eigenvalue of a Wishart matrix\,\cite{CHIANI_RMT} (see Eq. \ref{kappa_1_distribution} for an explicit expression) expresses the eigenvalue distribution of $W$ in terms of a scaled and shifted Gaussian distribution, itself derived as an approximation to the Tracy-Widom distribution \cite{TracyWidom} which is asymptotically exact for $n,m\rightarrow \infty$. Using this, we find that the normalised probability density function (PDF) of the largest singular value $\lambda_1$ defined for $y>0$ (or $y>\sqrt{\mu-\rho\alpha}$ if $\mu-\rho\alpha>0$,) is approximately given by 
\begin{equation}
    f_{\lambda_1}(y) = \frac{2}{\rho\theta^k \Gamma(k)}y\left(\frac{y^2-\mu}{\rho}+\alpha\right)^{k-1}e^{-\frac{1}{\theta}\left(\frac{y^2-\mu}{\rho}+\alpha\right)},
    \label{lambda1distribution}
\end{equation}
where $k=79.6595,\theta=0.101037$ and $\alpha=9.81961$ are fitting parameters chosen to match the first 3 moments of the approximated Gaussian distribution to the type-2 Tracy Widom distribution\,\cite{CHIANI_RMT}. The scaling and centering parameters of the Tracy Widom 2 distribution\,\cite{johnstone2006_high,MA_Twscaling} are given by 
\begin{equation}
\mu = (\sqrt{n}+\sqrt{m})^2, \rho = \sqrt{\mu}\left(\frac{1}{\sqrt{n}} + \frac{1}{\sqrt{m}}\right)^{1/3}.
\end{equation}

\begin{figure}[t]
    \centering
    \includegraphics[width=0.8\linewidth]{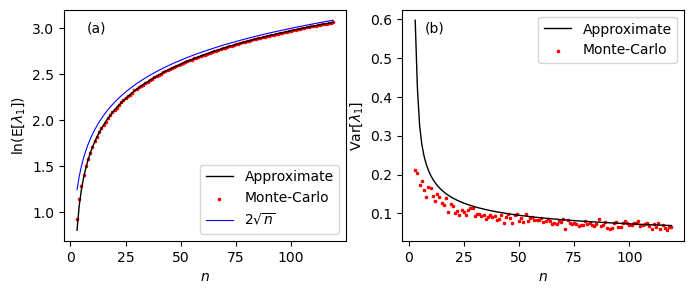}
    \caption{Approximate analytical estimates of (a) $\mathbb{E}[\lambda_1]$ (Eqs. \ref{expectedval} and \ref{2rt2}), and (b) $\mathrm{Var}[\lambda_1]$ (Eq. \ref{variance}), plotted against numerically calculated Monte-Carlo averages of these estimates from 1000 random systems for each $n$, with $n=m$. }
    \label{fig:MonteCarloEvVar}
\end{figure}

We find that the average largest singular value $\mathbb{E}[\lambda_1]$ can be approximated as \cite{abramowitz+stegun}:
\begin{equation}
    \mathbb{E}[\lambda_1] \approx \frac{\sqrt{\rho}}{\theta^k}\left(\frac{\mu}{\rho}-\alpha\right)^{k+1/2} \mathcal{U}\left(k,\frac{3}{2}+k,\frac{1}{\theta}\left(\frac{\mu}{\rho}-\alpha\right)\right),
    \label{expectedval}
\end{equation}
where $\mathcal{U}$ is Tricomi's confluent hypergeometric function. This expression turns out to be accurate even for small $m,n$, however in the asymptotic regime $n\rightarrow \infty$ with $m=n$, we recover the result that the average largest singular value scales as
\begin{equation}
    \mathbb{E}[\lambda_1] \sim 2 \sqrt{n}, \label{2rt2}
\end{equation}
which has also been derived directly via the Marchenko-Pastur distribution of the eigenvalues of large rectangular matrices\,\cite{marchenkopastur,bai_asymptotic2root2}.

Using Eq. \ref{lambda1distribution}, we find that the variance of the largest singular values is approximately given by
\begin{equation}
    \mathrm{Var}[\lambda_1] \approx \rho[\theta(k+1)-\alpha]+\mu - \mathbb{E}[\lambda_1]^2,
    \label{variance}
\end{equation}
which tends to be very small for large $n,m$, but otherwise does not have a simple asymptotic formula. Figures \ref{fig:MonteCarloEvVar}(a)(b) compare our approximate estimates for the average and variance of the largest singular value (Eqs.\ \ref{expectedval} and \ref{variance}) to a Monte-Carlo sampled average of these values over 1000 random systems as a function of $n(=m)$. The estimates derived do indeed work well, especially for the expected value. Indeed, Eq.\ \ref{expectedval} agrees with the simulations all the way down to $n=2$, representing a significant improvement over the asymptotic result, $2\sqrt{n}$. Our variance estimate is less good for small $n$, due to higher moments being more susceptible to compounding errors caused by successive approximations, but improves quickly as $n$ increases. The variance plotted in Fig.\ \ref{fig:MonteCarloEvVar}(b) confirms that, even for randomized coupling matrices, the system exhibits relatively predictable behaviour, with the variance decreasing as $n$ increases. This implies that for all but the very smallest systems, a stable coupling boost $\sim 2\sqrt{n}$ should be a fairly stable feature of samples taken from the ensemble of many-level QRM system with random couplings.

\begin{figure}[t]
    \centering
    \includegraphics[width=0.8\linewidth]{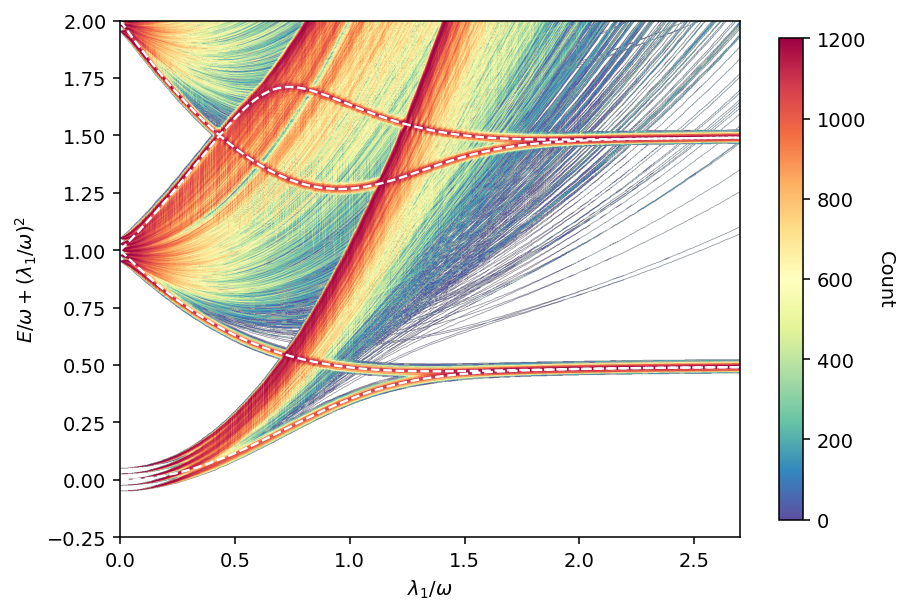}
    \caption{Density of the energy levels for 600 systems with $n=m=5$ and coupling matrices sampled from a random distribution, normalised so that the primary coupling is always 
    $\lambda_1$ (see the main text for further details). The `count' gives the density of energy levels within discrete bins ($1000\times1250$ in total) covering the range of energies and couplings. Energy levels of a simple QRM with coupling $\lambda = \lambda_1$ are indicated with white dotted lines for comparison. }
    \label{fig:heatmap}
\end{figure}

\subsection{Energy Level Distribution for Random couplings}
We now look more broadly at the distribution of the energy levels for randomly coupled systems, going  beyond just those controlled by the largest singular value. Figure \ref{fig:heatmap} maps the density of the energy levels from 600 systems with $n=m=5$ as a function of the largest coupling strength, $\lambda_1$. The systems have coupling matrices drawn from the random distribution $\Lambda_{ij}\sim \lambda_1\mathscr{CN}(0,1)/{\rm[Max_{SV}}\mathscr{CN}(0,1)]$, where here $\lambda_1$ is simply a scale factor varied to change the overall strength of the couplings and ${\rm[Max_{SV}}\mathscr{CN}(0,1)]$ is the maximum singular value of the corresponding sampling from the $\mathscr{CN}(0,1)$ distribution. Thus, by construction, the value of the largest singular value is equal to $\lambda_1$ for each of the 600 sampled systems. Weak detunings have also been included, $\varepsilon/\omega=0.05$, with the individual detunings of both the ground and excited levels equally spaced between $-0.05\omega$ and $0.05\omega$. 



Figure \ref{fig:heatmap} shows that with appropriate scaling, the overall spectral structure of a traditional QRM with coupling $\lambda_1$ is largely preserved within the spectrum of a multilevel system across many realisations of the random coupling matrices (the spectrum of the standard 2-level QRM with $\lambda_1=\lambda$ is shown with dashed lines for comparison). The highest count densities in the heatmap correspond to states exhibiting either the largest light-matter interaction (for which $\langle\hat{D}\rangle\approx1$), which align with the 2-level QRM energy levels, or very weak couplings (i.e. the `darkest' states). In the latter case, the energy eigenergies only increase extremely slowly as a function $\lambda_1$, leading to almost perfectly quadratic curves in  Fig.\ \ref{fig:heatmap} due to the additional factor of $(\lambda_1/\omega)^2$ added to each of the energies. 

\begin{figure}[t]
    \centering
    \includegraphics[width=0.6\linewidth]{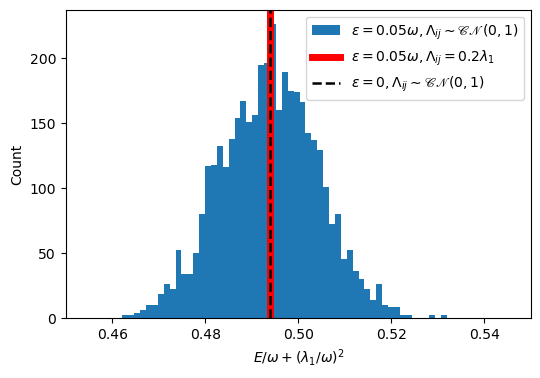}
    \caption{Histogram of energy levels of the two almost-degenerate groundstate energies of 600 random systems with the same parameters as those in Fig. \ref{fig:heatmap} and $\lambda_1/\omega=2.5$. Vertical lines showing the energies of similar systems with either no detunings ($\varepsilon=0$), or a uniform coupling matrix ($\Lambda_{ij} = 0.2\lambda_1$). Only the case with both nonzero detuning and non-uniform coupling matrices yields a significant variation in these energy levels ($\sim \varepsilon/\omega$). 
    }
    \label{fig:gstate_variation_histogram}
\end{figure}

Thus far we have seen that the behaviour of the most strongly coupled levels of a randomly coupled system can almost be mapped onto those of a standard 2-level QRM. However, when $\varepsilon\neq 0$, the mapping is imperfect and samples of even the most strongly coupled $(\langle D\rangle\simeq 1)$ states give rise to energy levels that are distributed over a finite range in Fig.\ \ref{fig:heatmap}. To see this more clearly, Fig.\ \ref{fig:gstate_variation_histogram} shows the distribution of the two lowest energy levels for $\lambda_1/\omega=2.5$.  

The behaviour of the two lowest lying energy levels (which have opposite parity) in the extreme coupling limit ($\lambda_1/\omega\rightarrow+\infty$) is well known in the standard QRM: $E/\omega+(\lambda_1/\omega)^2 \rightarrow 0.5$\,\cite{Rossatto2017,Twyeffort2007}. For the randomly coupled system with finite detunings, different samples lead to a distribution of the two lowest energy levels with finite variance. In Fig. \ref{fig:gstate_variation_histogram}, we compare the distribution of the two nearly degenerate ground state energies at $\lambda_1=2.5$ to cases with a constant coupling matrix $\Lambda_{ij}=\lambda_1/\sqrt{nm}=0.2\lambda_1$ (and finite detunings), or no detunings, $\varepsilon=0$, and random couplings. In the latter two situations, a tiny residual separation between the two levels ($\sim 10^{-12}\omega$) arises because the coupling, although very large, remains finite. The distribution of energy levels thus requires both a randomly coupled system and non-zero detunings ($\varepsilon\neq 0$). It reflects the mixing of doublets within the eigenstates (the corresponding value of $\langle D\rangle$ is not precisely unity) caused by finite detuning and hence they do not depend on just the largest singular value. Thus, scaling by $\lambda_1$ cannot fully remove the randomness of the underlying coupling matrix leading to a distribution of finite width.

\section{Conclusions}
\label{sec:CCs}
We have explored the properties of a generalisation of the celebrated QRM in which the two-level atom is replaced by two manifolds of atomic levels with the separation between levels within each manifold much smaller than the average separation between the manifolds. We found that when the levels within each manifold are actually degenerate, the system resolves into a direct sum of two-level QRMs. However, the couplings within each of the QRM are in general different and depend strongly on the form of the original coupling matrix between the individual levels in the two manifolds, together with the total number of the states. Most significantly, the strongest coupling is enhanced in a way that scales with $n$ for uniform couplings. For random couplings, we derived an approximate expression for the behaviour of the strongest coupling [Eq.\ \ref{expectedval}] which reduces to $2\sqrt{n}$ for large enough $n$. We also found that although the broad structure of the spectral properties are not significantly changed by the introduction of small detunings, closer examination reveals that they can nevertheless lead to interesting systematic deviations.

We conclude that multilevel systems in which the individual elements of the coupling matrix are relatively weak, can nevertheless provide a promising route to accessing the strong coupling regimes of the Rabi model, provided sufficiently large numbers of almost degenerate levels exist. Going beyond the case of multiple excited states considered in Ref.\ \cite{Tufarelli_2021}, having in addition a ground state manifold containing a large number of almost degenerate levels does indeed provide a significant additional boost to the largest coupling in the system. Furthermore, our results from the extreme cases of uniform and random coupling matrices suggest that the strongest coupling for the more likely range of intermediate cases will be boosted in a way that scales at least as strongly as $2\sqrt{n}$. 

We believe that there are several interesting possible ways in which the work presented here could be extended. These include quantifying the impact of detunings on the spectral properties of the system in more detail (e.g., calculating the properties of the distribution shown in Fig.\ \ref{fig:gstate_variation_histogram} analytically), as well exploring how they affect the dynamical properties.  
Finally, we note that in thinking about the couplings between levels in the ground- and excited-state manifolds, we have approached the problem from a mathematical perspective focusing on the limiting cases of uniform and random couplings. For specific physical realisations (colloidal quantum dots coupled to a plasmonic resonator\,\cite{Gross2018}, for example) it would be interesting to carry out a derivation starting from a microscopic description. This might be expected to reveal system-dependent constraints on the couplings, as well as possibly giving rise to interesting complications, such as couplings to more than one field quadrature.

\section*{Acknowledgments}
We thank Sven Gnutzmann for very helpful suggestions on random matrix theory. ADA acknowledges support in the form of a Leverhulme Trust Research Project Grant (RPG-2023-177).

\appendix

\section{Singular Value Decomposition and Radiation Basis}
\label{app:SVD}
The goal of this Appendix is to outline the general form of the transformation between the bare basis and radiation basis achieved by applying a singular value decomposition (SVD) to the coupling matrix, $\bL$. The SVD isolates the collective atomic modes that couple independently to the field, with each singular value $\lambda_k$ representing an effective coupling strength. This decomposition ensures that the light-matter interaction is reduced to a direct sum of independent two-level subsystems. Only those modes with non-zero singular values, of which there are at most $m$ (since we take $n\ge m$), participate in the dynamics when $\varepsilon=0$. 

We start with the atomic part of the light-matter interaction part of the multilevel QRM Hamiltonian [Eq. \ref{MQRMhamiltonian}]
\begin{equation}
    \hat{H}_{int}^{(a)} = \sum_{j=1}^m\sum_{i=1}^n\left(\Lambda_{ij}\sij{ji}+\Lambda_{ij}^*\sij{ji}^\dagger\right).
\end{equation}
In general, the numbers of ground and excited levels will not be the same ($n\neq m$), so that $\bL$ is a rectangular matrix, and so we need to use a SVD, where we decompose via
\begin{equation}
\boldsymbol{\Lambda}=\bU^\dagger\boldsymbol{\lambda}\boldsymbol{V},
\end{equation}
where $\boldsymbol{\lambda}$ is a pseudo-diagonal matrix encoding the singular values of $\bL$, while $\bU$ and $\bV$ are unitary matrices. Carrying out this decomposition (again assuming $n\ge m$) yields
\begin{equation}
\hat{H}_{int}^{(a)} = \sum_{j,i=1}^{m,n} \sum_{k=1}^{m} \left[\left(U_{ik} \lambda_k V^\top_{kj}\right) \ket{g_i}\bra{e_j} + \left(U_{ik} \sigma_k V^{\top}_{kj}\right)^*\ket{e_j}\bra{g_i}\right].
\end{equation}
We define the radiation basis states as
\begin{equation}
    \ket{G_k} = \sum_{i=1}^{m} U_{ik} \ket{g_i}, 
 \ \mathrm{and} \ \ket{E_k} = \sum_{j=1}^{n} V^\top_{kj} \ket{e_j}
\end{equation}
with
\begin{equation}
    \Sij{ji}\equiv\ket{G_j}\bra{E_i},
\end{equation}
so that in the new basis the light-matter interaction term can be more simply expressed as
\begin{equation}
   \hat{H}_{int}^{(a)} = \sum_{k=1}^{m}\lambda_k (\ket{G_k}\bra{E_k} + \ket{E_k}\bra{G_k}).
   \label{Haminteractionappendix}
\end{equation}

The radiation basis is orthonormal, thanks to the unitarity of $\bU$ and $\bV$. Additionally, we can see from the construction of Eq. \ref{Haminteractionappendix} that any excess excited states $\ket{E_k}$ where $k>m$ are automatically excluded from the light-matter interaction. These states can only couple to others through the detuning terms and are therefore automatically dark states. We can also define a more general doublet type operator in the radiation basis:   
\begin{equation}
    \hat D =\sum_{k=1}^m k\left(\proj{E_k}+\proj{G_k}\right). \label{eq:svdtype}
\end{equation}

\section{Statistics of Largest Singular Values of the Real and Complex Uncorrelated Ginibre Ensemble}
\label{app:stats}
In this Appendix we provide details on the derivations of the results discussed in Sec.\ \ref{random couplings section} of the main text, relating to the approximate statistics of the largest singular values of random Ginibre matrices of arbitrary size $n\times m$. We start with a random Ginibre matrix $\Lambda$, with IID Gaussian (real or complex) entries of unit variance $\Lambda_{ij} \sim \mathscr{N}(0,1)$. It is known\,\cite{CHIANI_RMT} that the probability distribution function (PDF) of the distribution of the largest eigenvalue of the Wishart matrix given by $W = \bL \bL^T$, $\kappa_1$, is approximately described by an appropriately scaled and shifted Gamma distribution valid for $z>\mu-\alpha\rho=\min{(\kappa_1)}$,
\begin{equation}
    f_{\kappa_1}(z) = \frac{1}{\rho \theta^k \Gamma(k)}\left(\frac{z-\mu}{\rho}+\alpha\right)^{k-1}e^{-\frac{1}{\theta}\left(\frac{z-\mu}{\rho}+\alpha\right)},
    \label{kappa_1_distribution}
\end{equation}
itself an approximation of a Tracy-Widom distribution\,\cite{TracyWidom}, which is asymptotically exact for $n,m\rightarrow\infty$. $\mu$ and $\rho$ are chosen to approximate the true distribution by matching it to the appropriate Tracy-Widom distribution. They are given by\,\cite{johnstone2006_high,MA_Twscaling}
\begin{equation}
\mu = (\sqrt{n+a_1}+\sqrt{m+a_2})^2, \rho = \sqrt{\mu}\left(\frac{1}{\sqrt{n+a_1}} + \frac{1}{\sqrt{m+a_2}}\right)^{1/3}.
\end{equation}

Table \ref{table} provides the parameter values for the complex and real cases. The parameter values $k,\theta,\alpha$ are taken from Ref.\ \cite{CHIANI_RMT} and match the relevant Tracy-Widom distribution to the Gamma distribution. The parameters $a_1$ and $a_2$ serve as finite-size corrections in the Tracy-Widom scaling\,\cite{johnstone2006_high,MA_Twscaling}.

\begin{table}[t]
\begin{center}
\begin{tabular}{ |p{3.5cm}||p{1.5cm}|p{1.5cm}|p{1.5cm}|p{1.5cm}|p{1.5cm}|  }
 \hline
 \multicolumn{6}{|c|}{Parameter Table} \\
 \hline
 
 Matrix Type& $k$ & $\theta$ & $\alpha$ & $a_1$ & $a_2$\\
 \hline
 \hline
 Real Ginibre   & $46.446$    &$0.186054$&   $9.84801$& $-1/2$& $-1/2$\\
 Complex Ginibre&   $79.6595$  & $0.101037$   &$9.81961$&$0$ &$0$\\
 \hline
\end{tabular}
\end{center}
\caption{Table of parameter values needed for Eq.\ \ref{kappa_1_distribution}.}\label{table}
\end{table}

Since the singular values of a matrix $\bL$ correspond to the square roots of the eigenvalues of the associated Wishart matrix $W = \bL\bL^T$, the largest singular value is given by $\lambda_1 = \sqrt{\kappa_1}$.
It is then straightforward to carry out a transformation of random variables to find the PDF of the largest singular value $\lambda_1$, which corresponds to the shifted exponential Tricomi distribution\,\cite{fitz_tricomi_dist}
\begin{equation}
    f_{\lambda_1}(y) = \frac{2}{\rho\theta^k \Gamma(k)}y\left(\frac{y^2-\mu}{\rho}+\alpha\right)^{k-1}e^{-\frac{1}{\theta}\left(\frac{y^2-\mu}{\rho}+\alpha\right)},
    \label{lambda1distributionappendix}
\end{equation}
which we take to be valid for $y>\sqrt{\min(\kappa_1})$. There is some subtlety here, as strictly speaking we can have $\min(\kappa_1)<0$ so $y$ is implied to enter the imaginary domain for some choices of $n,m$. Whenever this occurs we simply require $y>0$ instead. Although somewhat ad-hoc, this procedure can be justified on the basis that singular values cannot be imaginary. Although the case $\min(\kappa_1)<0$ is cumbersome to deal with when calculating moments analytically, due to the integral limits not being amenable to analytical solutions, evaluating the expressions despite this yields good results and it is possible to verify the PDF is still a unit measure numerically.

\begin{figure}[t]
    \centering
    \includegraphics[width=0.49\linewidth]{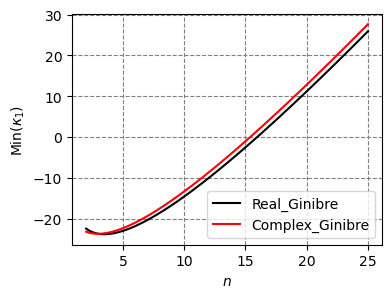}
    \caption{Minimum value of $\kappa_1$ in the Gamma distribution, Eq.\ \ref{kappa_1_distribution}, as a function of $n$ (with $n=m$) for the real and complex cases. The values become negative when $n\leq16$.}
    \label{fig:minkappa}
\end{figure}

\begin{figure}[t]
    \centering
    \includegraphics[width=0.95\linewidth]{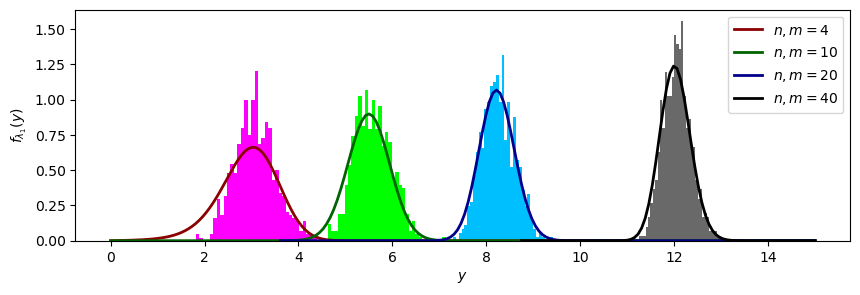}
    \caption{Derived largest singular value distributions (lines) [Eq.\ \ref{lambda1distributionappendix}] compared to empirically calculated histograms of largest singular values for 600 randomly sampled matrices. The colours indicate different matrix sizes. It is clear that the approximate PDF $f_{\lambda_1}(y)$ is quite accurate, even for small $n(=m)$, and errors that do occur (noticeably around the left tail of the distribution) dissipate rapidly with increased matrix size.}
    \label{fig:approx_distributions}
\end{figure}

When $n=m$, this translates into our results being rigorously analytically correct when $n>16$ in both complex and real cases. In Fig. \ref{fig:minkappa}, we plot $\min{(\kappa_1)}$ as a function of $n$, where $n=m$. The figure shows the minimum possible value of $z$ for the eigenvalue distribution, Eq.\  \ref{kappa_1_distribution}. Note that much of the left tail of $f_{\kappa_1}(z)$ is near $0$ for most cases due to the low overall variance, and that indeed in theory we must have $\kappa_1,\lambda_1>0$ as a result of the construction of the Wishart matrix and singular values. The fact that $f_{\kappa_1}(z)$ can be defined for $z<0$, or that $f_{\lambda_1}(y)$ could be interpreted as being defined for imaginary $y$ for $n\leq16$ can be seen as a compounding error caused by successive approximations. In Fig. \ref{fig:approx_distributions}, we compare our derived distribution for the largest singular value, $f_{\lambda_1}(y)$, against Monte Carlo simulations of the distribution for various $n$ values (with $n=m$ in each case), each with 600 trials. We can see that our approximate distribution is quite good even for small matrix sizes and quickly improves as $n$ becomes larger. This follows from the asymptotic validity of Tracy-Widom scaling, which ensures that our approximation becomes more accurate in the large $n,m$ limit.

We can also derive expressions for the moments of our largest singular value\,\cite{abramowitz+stegun}  
\begin{equation}
\fl
    \mathbb{E}[\lambda_1^l] = \int_{\sqrt{\mu-\alpha\rho}}^{\infty} y^l f_{\lambda_1}(y)\d x = \frac{\rho^{\frac{l}{2}}}{\theta^k}\left(\frac{\mu}{\rho}-\alpha\right)^{\frac{l}{2}+k}\mathcal{U}\left(k,\frac{l}{2}+k+1,\frac{1}{\theta}\left(\frac{\mu}{\rho}-\alpha\right)\right),
\end{equation}
where $\mathcal{U}$ is Tricomi's confluent hypergeometric function. When $l=0$, and using the identity $\mathcal{U}(a,a+1,x) = x^{-a}$ \cite{abramowitz+stegun}, we can confirm that our distribution is a unit measure. For the mean ($l=1$), we use that $\mathcal{U}(a,b,x)\sim x^{-a}$ \cite{abramowitz+stegun} asymptotically as $x\rightarrow \infty$. In our case, this would correspond to $n,m\rightarrow\infty$, and so
\begin{equation}
    \mathbb{E}[\lambda_1] \sim \sqrt{\mu-\alpha\rho} = 2^{2/3}n^{1/6}\sqrt{2^{2/3}n^{2/3}-\alpha} \sim 2\sqrt{n},
\end{equation}
for $n\rightarrow \infty$. For whenever $\min(\kappa_1)<0$, we verified numerically in Fig. \ref{fig:MonteCarloEvVar} that the expressions \ref{expectedval} and \ref{variance} still yield accurate results for the mean and variance when using the principal branch of the Tricomi function\,\cite{abramowitz+stegun}. Note that the principal branch is used automatically when no specific determination is made in the SciPy package. It is also important to emphasise that when numerically calculating either the PDF of the distribution itself or its moments, the terms being multiplied together in the expressions are very varied in magnitude, leading to significant errors in computation. This sometimes requires the extra step of using arbitrary floating point precision arithmetic.

In this Appendix we have derived a novel approximate distribution for the largest singular value of uncorrelated Ginibre matrices using an approximation for the largest eigenvalue of Wishart matrices\,\cite{CHIANI_RMT}. We have also validated that errors are small when using the derived approximation. This is true even for the small values of $n,m$, and the accuracy improves quickly as the size of the matrix grows, as left tail errors in $f_{\kappa_1}$ (see Fig.\ \ref{fig:approx_distributions}), and subsequently $f_{\lambda_1}$, shrink rapidly. Our approximations in this Appendix yield a novel and computationally easier way to calculate accurate singular value statistics for real and complex Ginibre ensemble matrices.
\section*{References}
\bibliographystyle{iopart-num}
\bibliography{MultilevelPaper}

\end{document}